\documentclass{ptapap}

\author{Richard I. Anderson}[ESO]
\affil[ESO]{European Southern Observatory\\
Karl-Schwarzschild-Str.\ 2, D-85748 Garching b. M\"unchen, Germany}

\title{Radial Velocity Observations of Classical Pulsating Stars}

\begin{document}

\maketitle

\begin{abstract}

I concisely review the history, applications, and recent developments pertaining to radial velocity (RV) observations of classical pulsating stars. The focus lies on type-I (classical) Cepheids, although the historical overview and most technical aspects are also relevant for RR~Lyrae stars and type-II Cepheids. The presence and impact of velocity gradients and different experimental setups on measured RV variability curves are discussed in some detail. Among the recent developments, modulated spectral line variability results in modulated RV curves that represent an issue for Baade-Wesselink-type distances and the detectability of spectroscopic companions, as well as challenges for stellar models. Spectral line asymmetry (e.g. via the bisector inverse span; BIS) provides a useful tool for identifying modulated spectral variability due to perturbations of velocity gradients. The imminent increase in number of pulsating stars observed with time series RVs by {\it Gaia} and ever increasing RV zero-point stability hold great promise for high-precision velocimetry to continue to provide new insights into stellar pulsations and their interactions with atmospheres.
\end{abstract}

\section{Introduction}

Classical pulsating stars such as RR~Lyrae stars and type-I \& II Cepheids feature radial pulsations that manifest as periodic spectral variability.
Radial velocity (RV) observations reduce the complexity of spectral variability (in $T_{\rm{eff}}$, $\log{g}$, turbulence, velocity gradients, line asymmetries, among others) to a series of easy-to-interpret Doppler shifts in units of $\rm{km\,s^{-1}}$. Although RVs are convenient quantities to work with, their seeming simplicity can be misleading. This review showcases some subtleties concerning pulsating star RVs, starting with a historical perspective.

Optical spectroscopy and velocity measurements of Cepheids were pioneered by \citet{1895ApJ.....1..160B}. Cepheid spectra provided crucial information for establishing the {\it pulsation hypothesis} \citep{1918MNRAS..78..639L,1926AN....228..359B,1946BAN....10...91W} and helping to distinguish between RR~Lyrae stars and Cepheid variables \citep{1916ApJ....44..273S}. Adding to the confusion, \citet{1929PASP...41...56M} correctly determined the $30\,$yr orbital period of Polaris. With improved spectral resolution, \citet{1949PASP...61..135S} and \citet{1956PASP...68..137K} noted line doubling first in type-II and then in classical Cepheids. \citet{1928ApJ....67..319S,1949ApJ...109..208S} further pointed out that different line species feature different amplitudes, a fact that has received repeated attention also in the case of classical Cepheids \citep[e.g.][]{1972PASP...84..656W,1990ApJ...362..333S,1993ApJ...415..323B,2005MNRAS.362.1167P}. Line asymmetries accompanying the pulsations were first noted by \citet{1989ApJ...337L..29S} and continue to receive attention \citep{1992MNRAS.259..474W,2006A&A...453..309N,2016MNRAS.463.1707A,2017arXiv171105728B}. Given their common origin in the pulsations, it is likely that line doubling and line asymmetries relate to the same phenomenon. Of course, both effects complicate the very definition of an RV measurement.

Cepheid RV curves -- just like their light curves -- are well described by a Fourier series whose complexity depends on the pulsation period, $P_{\rm p}$, and mode. The behavior of Cepheid RV Fourier parameters for short-period Cepheids has been shown in detail by \citet{1990ApJ...351..606K}, and for long-period Cepheids by \citet{2016ApJS..226...18A}.

Detections of (single-lined) spectroscopic binaries (SBs) with multi-year orbital periods, $P_{\rm o}$, became  increasingly frequent in the late 1950s \citep{1959ApJ...130..769A}. Indeed, RVs turned out to be the most reliable method of determining the presence of companion stars \citep{2003IBVS.5394....1S}, and systematic studies focusing both on the detection and upper limits of companions are now possible thanks to long temporal baselines and high precision paired with stable instrumental zero-points \citep{2015AJ....150...13E,2016ApJS..226...18A}. While SBs are very frequent among classical Cepheids, this is not true for RR~Lyrae stars or type-II Cepheids. Indeed, the multiplicity of the only known SB type-II Cepheid was discovered thanks to eclipses \citep{2017ApJ...842..110P}. Despite long-term efforts to detect SBs among Cepheids, new cases are still being reported. In particular, low-mass companions around well-known bright Cepheids \citep[such as the prototype $\delta$ Cephei, cf.][]{2015ApJ...804..144A} and companions of Cepheids fainter than $\sim 8$th magnitude can now be discovered thanks to improved RV precision and instrument efficiency. Identifying Cepheids exhibiting orbital motion further provides strong support for achieving high parallax accuracy \citep{2016ApJS..226...18A}, e.g.\ with {\it Gaia} or the {\it Hubble Space Telescope} \citep{2016A&A...595A...1G,2016ApJ...825...11C}. As such, RV observations play an important role for anchoring the extragalactic distance scale trigonometrically. 

RV observations are also of direct importance to distance measurements via Baade-Wesselink type methods, cf.\ Kervella et al.\ {\it (this volume)}, which compare the total angular diameter variation to the linear radius variation ($\Delta R_{\rm{lin}}$) to determine a quasi-geometric distance. $\Delta R_{\rm{lin}}$ is computed by integrating the RV curve over phase and multiplying by a projection factor, $p$, which translates between the observed line-of-sight component of the radial pulsation and the true pulsational velocity. In so doing, many astrophysical details are reduced to a convenient number, $p$, which incorporates a high degree of astrophysical complexity and has been the focus of much research for several decades \citep{1982A&A...109..258B,1994ApJ...432..367S,2007A&A...471..661N,2016A&A...587A.117B}. 

Most aforementioned applications of Cepheid RV observations were feasible with instrumental precision of $\sim 300\,\rm{m\,s^{-1}}$. However, recent observations based on spectrographs featuring highly stable wavelength zero-points and contemporaneous wavelength drift corrections have revealed a new level of complexity intrinsic to Cepheid atmospheres \citep{2014A&A...566L..10A,2016MNRAS.463.1707A}. Indeed, such observations have shown that harnessing the extreme precision afforded by state-of-the-art RV instruments requires a thorough understanding of a) how RV measurements are obtained and defined, and b) the atmospheric physics of pulsating stars. The following sections are thus intended to clarify typical techniques used for measuring RVs (\S\ref{sec:method}) and to showcase some examples of the additional complexity observed using high-precision RV observations of Cepheids (\S\ref{sec:modulations}). Conclusions are presented in the final section \S\ref{sec:recommendations}.

\section{How to Measure RVs and Alienate People}
\label{sec:method}

The first large Cepheid RV catalog was created by \citet[$>100$ objects]{1937ApJ....86..363J} and followed by many others \citep[e.g.][]	{1955MNRAS.115..363S,1985SAAOC...9....5C,1988ApJS...66...43B,1994A&AS..105..165P,1999A&AS..140...79I,2002yCat.3229....0G}. RV here usually refers to a measurement based on {\it metallic lines} \citep{1949ApJ...109..208S}. The common underlying assumption is that metallic lines -- notably Fe\,I, Fe\,II, Ti\,II, Sc\,II, Sr\,II, Ca\,I, and Mg\,II -- move in phase and with identical amplitudes. Conversely, other lines -- e.g.\ $H_\gamma$, Ca\,II $H$ and $K$ -- move at 	different velocities and are shifted in phase. 

\begin{figure}
\centering
\includegraphics{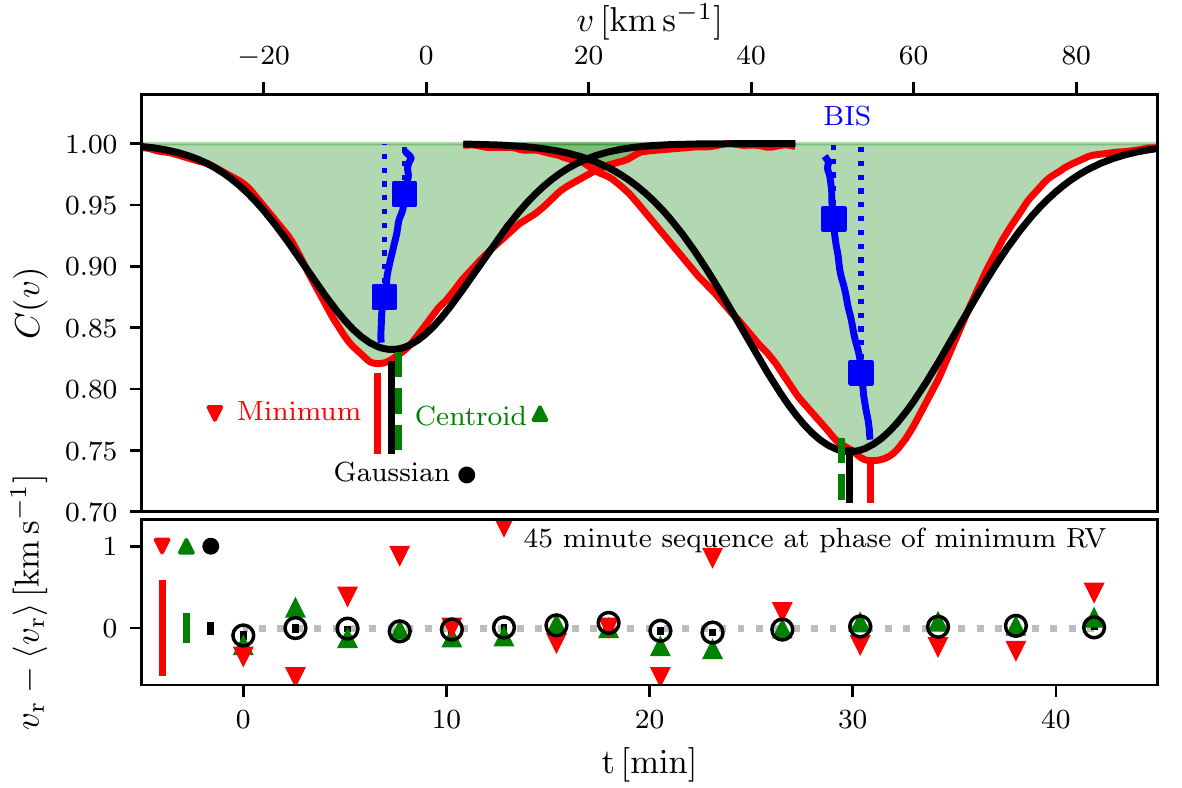}
\caption{{\it CORALIE} observations of the $20$\,d Cepheid RZ\,Vel \citep{2013PhDT.......363A}. {\it Top:} Illustration of different RV measurement definitions and the bisector inverse span (BIS), cf.\ \citet{2016MNRAS.463.1707A}, based on observed CCFs (red lines) near minimum and maximum RV. Fitted Gaussians are shown as black lines. {\it Bottom:} RV precision as function of RV definition illustrated using a 45-minute sequence of observations near minimum RV, where no RV variation is expected. Gaussian RVs (black circles) are the most precise and exhibit the smallest scatter (RMS shown as vertical bars to the left).  Centroid RVs are shown as green upward triangles, minimum CCF RVs as red downward triangles. Line minimum RVs are the least precise and yield the largest peak-to-peak amplitude.}
\label{fig:RVdefinition}
\end{figure}

When combining data from different sources, one must therefore consider differences in experimental setup as well as effects related to stellar astrophysics and changing instrumental properties. For instance: different spectral lines exhibit different velocities, even metallic lines (\S\ref{sec:modulations}); non-linear changes in pulsation periods 
% \citep[e.g.][]{2017arXiv170310334S} 
complicate RV curve de-phasing and measuring the pulsation-subtracted velocity, $v_\gamma$; seasonal changes in temperature and ambient pressure lead to changing instrument zero-points \citep{2015AJ....150...13E}.

There are different RV measurement definitions that each have their own advantages and disadvantages. Definitions used in the literature include: a) Gaussian RVs, b) Centroid (or line barycenter) RVs, c) Line minimum RVs, and d) bi-Gaussian RVs, among others. Importantly, any RV measurement is biased with respect to line asymmetry, and therefore, to RV amplitude. Whereas centroid RVs are often said to be the most physical (because they represent a weighted average of the line profile), a key benefit of using Gaussian RVs is their insensitivity to noisy line profiles that renders Gaussian RVs particularly precise and easy to measure. Line minima -- although the eye is drawn to them -- carry the least RV information, since the velocity derivative is 0 at a line's core by definition. Finally, bi-Gaussian RVs (poorly) account for line asymmetry, and lead to systematically larger pulsation amplitudes than Gaussian RVs. Such subtle differences significantly influence any studies interested in values of $v_\gamma$ or pulsation amplitudes. Figure\,\ref{fig:RVdefinition} illustrates these differences and demonstrates that Gaussian RVs provide the most precise measurement, i.e., reproduce a stable mean RV  with the smallest scatter. The following discussion is based on Gaussian RVs, which benefit the most from state-of-the-art instrumental precision and stability.

\begin{figure}
\centering
\includegraphics{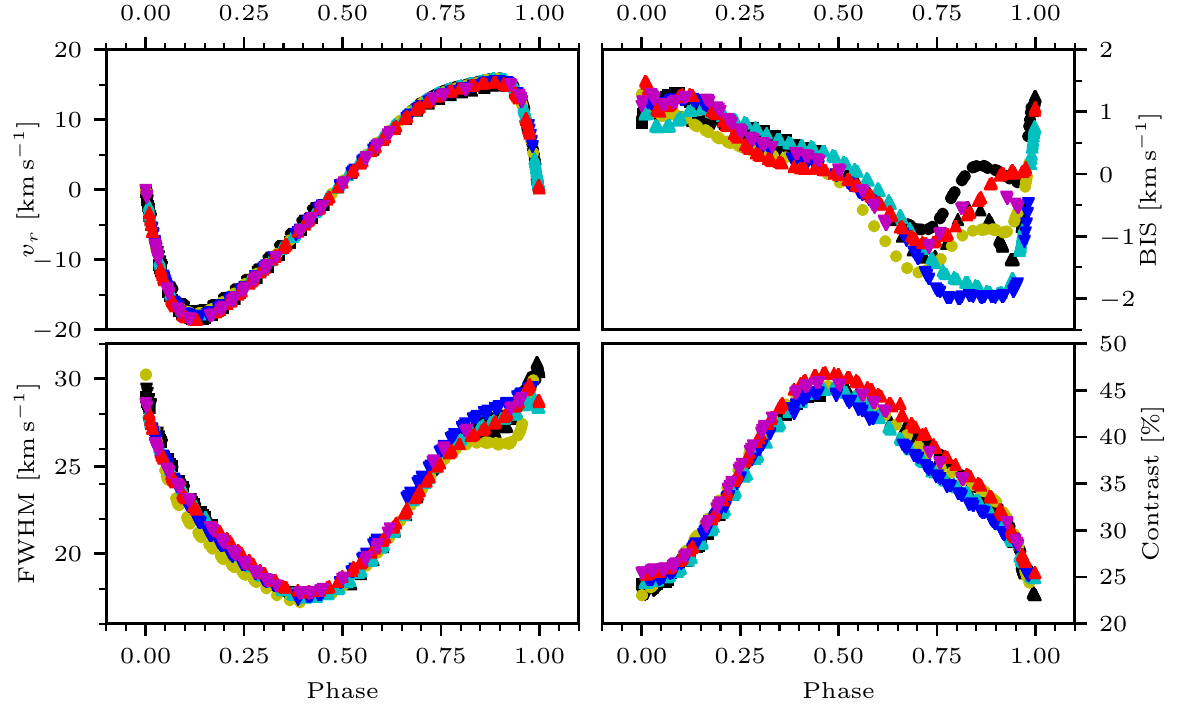}
\caption{Cycle-to-cycle modulated spectral variability seen in RV, BIS, FWHM, and CCF. Each shade represents a different pulsation cycle, cf.\ \citet[$\ell$ Car; $P_{\rm{p}} = 35$\,d]{2016MNRAS.463.1707A}.}
\label{fig:lCarCCFparameters} 
\end{figure}

The cross-correlation technique \citep{1996A&AS..119..373B,2002A&A...388..632P} awards significantly higher RV precision than measurements based on individual spectral lines. A {\it line mask}, composed of the positions and relative strengths of thousands of spectral lines is successively shifted in velocity space and multiplied with the observed stellar spectrum. The results are smooth cross-correlation functions (CCFs, Fig.\,\ref{fig:RVdefinition}) that feature a $\sim 60$-fold increase in signal-to-noise ratio (SNR) compared to individual spectral lines. Thus, CCFs enable precise RV measurements of even relatively faint Cepheids for which a typical SNR might be on the order of $10$ per pixel. Of course, the choice of line mask directly impacts the shape of the resulting CCF and forms part of the RV zero-point. In the case of pulsating stars, where atmospheric layers are moving differentially, this means that line masks (through selection and weighting by line strength) introduce a weighting of different atmospheric depths. Moreover, this weighting differs between Cepheids of different average spectral types, and changes as a function of pulsation phase for every Cepheid.

CCFs contain much more information about a Cepheid's atmosphere than just a line-of-sight velocity. Line contrast (depth) and full width at half-maximum (FWHM) are often derived for the Gaussian profile fitted to the CCF and contain information about the variability of temperature (spectral type) and turbulence. The Bisector Inverse Span \citep[BIS, cf.][]{2001A&A...379..279Q} is derived from the CCF itself and quantifies CCF asymmetry, see Fig.\,\ref{fig:RVdefinition}. BIS is defined as the velocity difference between the average CCF bisector velocities measured near the top and the bottom of the CCF, and exhibits a characteristic phase-dependent variability that is particularly sensitive to velocity gradients. Figure\,\ref{fig:lCarCCFparameters} shows the four parameters RV, BIS, FWHM, and contrast in the case of the $35$\,d Cepheid $\ell$~Carinae \citep{2016MNRAS.463.1707A}.

Different RV measurement definitions can be applied to computed CCFs as well as individual spectral lines, resulting in a large number of possible experimental setups (combinations of lines used; RV definition applied). For instance, in the case of double-lined eclipsing binaries, the broadening function method may yield better results than CCFs \citep{1992AJ....104.1968R,2017ApJ...842..110P}. It is thus clear that authors who discuss and publish pulsating star RVs should take care to unambiguously describe exactly how RV was derived. Better yet, authors should publish not just the derived RVs, but include reduced spectra, line masks used for computation, and computed CCFs to ensure transparency and reproducibility. Those interested in data from {\it Gaia}'s RVS instrument should be aware of the different methodology used to measure RV, which may lead to zero-point and amplitude differences between the calcium IR-triplet lines and optical metallic lines (Anderson, Wallerstein, et al. in prep.).

\section{High Precision Velocimetry Reveals Modulated RV Variability}
\label{sec:modulations}

High-cadence, high-quality RV observations using modern (wavelength) stabilized high-resolution spectrographs have enabled the recent discovery that Cepheid RV curves exhibit small modulations, which differ in timescale and amplitude between short-period overtone pulsators (V335~Puppis, QZ~Normae) and long-period ($P_{\rm{p}} > 20$\,d, $\ell$~Carinae and RS~Puppis) Cepheids \citep{2014A&A...566L..10A}. Additional cases were recently presented by \citet{2016ApJS..226...18A}. Notably, RV curve modulations in overtone pulsators have long ($\sim$ years) timescales with as yet no detected periodicity. Conversely, RV curve shapes differ between subsequent pulsation cycles in long-period Cepheids, i.e., every pulsation cycle is different. Notably, line asymmetry (BIS) is an excellent indicator for the presence of cycle-to-cycle variations (cf.\ Fig.\,\ref{fig:lCarCCFparameters}). 

RV curve modulation represents an issue for many Cepheid-related studies (e.g.\ Baade-Wesselink-type methods), since changing amplitudes can result in $\sim 5 - 15\%$ systematic differences in $\Delta R_{\rm{lin}}$ or $p-$factors \citep{2014A&A...566L..10A}. \citet{2016MNRAS.455.4231A} have further shown that contemporaneous observations of angular and linear radius variations do not resolve the issue, since the variability of angular diameters appears to be modulated differently than the variation of linear radii. Moreover, RV curve modulation is generally not symmetric around the mean, so that $v_\gamma$ inferred from different cycles can appear to fluctuate. RV curve modulation hence limits the detectability of spectroscopic companions via time-variable $v_\gamma$ and introduces a time-dependence in the $k-$term problem \citep{1937ApJ....86..363J,2009A&A...502..951N}. 

\begin{figure}
\centering
\includegraphics{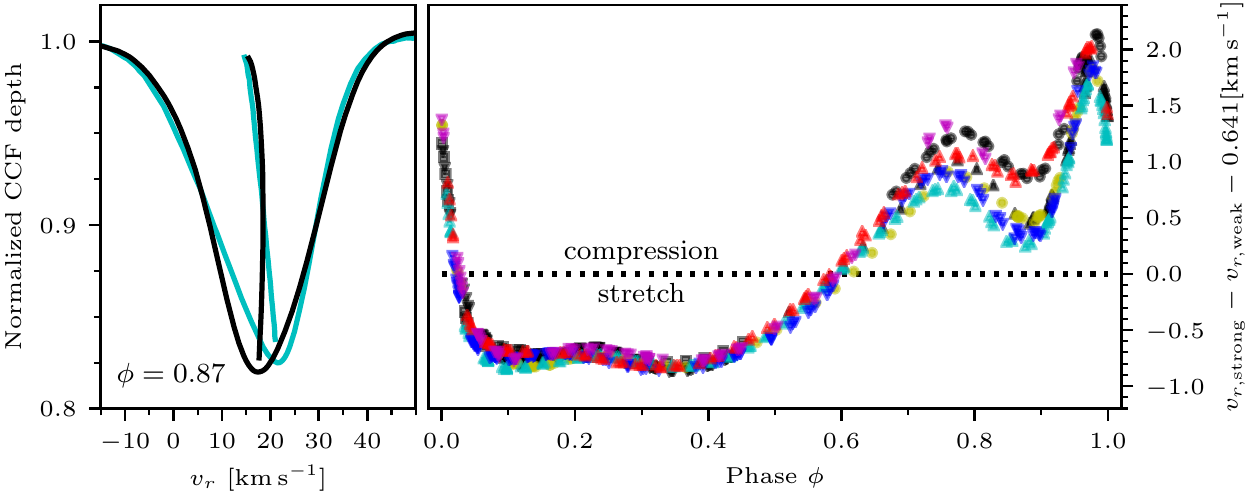}
\caption{Cycle-dependent CCF profiles at phase $0.87$ (left) due to metallic line velocity gradient perturbations visualized using Doppler tomography (right), see \citet{2016MNRAS.463.1707A}.}
\label{fig:CCFmodulation}
\end{figure}

Cycle-to-cycle variations in long-period Cepheids originate from velocity field perturbations that act on timescales longer than $P_{\rm{p}}$ and carry information between pulsation cycles \citep{2016MNRAS.463.1707A}. These velocity field perturbations lead to different line shapes observed at identical pulsation phases during different cycles, cf.\ Fig.\,\ref{fig:CCFmodulation}. This discovery was made by visualizing atmospheric velocity gradients among metallic lines using a Doppler tomographic technique that used two correlation masks composed of exclusively weak and strong lines to measure RV in two different atmospheric layers. Thus, Doppler Tomography revealed the phase dependence of metallic line velocity gradients as well as the pulsation cycle-dependence of their variability. Several possible explanations for this behavior exist \citep[cf.][and references therein]{2016MNRAS.463.1707A}. While no definitive answer is as yet available, an interaction between the pulsation and convection does appear as a likely candidate given the large convective cells in low-density atmospheres of cool long-period Cepheids. 

\section{Conclusions}
\label{conclusions} \label{sec:recommendations}

RV observations of classical pulsators have played an important role for  understanding stellar pulsations and calibrating the cosmic distance scale. 
{\it Gaia}'s RVS instrument measures RV time-series on the order of $10^4 - 10^5$ pulsating stars. Authors interested in combining heterogeneous RV data (e.g.\ {\it Gaia} RV data with other literature) must take care to understand, and possibly correct for, differences in zero-points and amplitudes due to heterogeneous measurement definitions (e.g.\ Gaussians or centroids) and velocity gradients (different lines yield different RVs). Conversely, authors who publish RV data should be mindful of such issues and aim for maximum transparency and reproducibility.

Modern stabilized high-resolution spectrographs afford instrumental RV precision sufficient to reveal new pulsation-related phenomena, such as cycle-to-cycle and long-term spectral modulations. Understanding and learning to mitigate these effects are crucial for improving the precision of Baade-Wesselink distances and the ability to detect low-mass stellar companions. Investigating the relation between modulated photometric \citep[e.g.][]{2015MNRAS.446.4008E} and RV variability will provide further insights into the complex interplay of pulsations and stellar atmospheres.

\acknowledgements{It is my pleasure to acknowledge useful discussions with the Geneva exoplanet group, especially with Xavier Dumusque, Francesco Pepe, and Christophe Lovis. Thanks are also due to the support teams of the Euler and Mercator telescopes.}

\bibliographystyle{ptapap}
\bibliography{RIAbib}

\begin{thebibliography}{48}
\providecommand{\natexlab}[1]{#1}
\providecommand{\url}[1]{\texttt{#1}}
\providecommand{\urlprefix}{URL }
\providecommand{\eprint}[2][]{\url{#2}}

\bibitem[{{Abt}(1959)}]{1959ApJ...130..769A}
{Abt}, H.~A., \emph{{The Cepheid Binary FF Aquilae.}}, \emph{\apj}
  \textbf{130}, 769 (1959)

\bibitem[{{Anderson}(2013)}]{2013PhDT.......363A}
{Anderson}, R.~I., {Classical Cepheids: High-precision Velocimetry, Cluster
  Membership, and the Effect of Rotation}, Ph.D. thesis, Universit{\'e} de
  Gen{\`e}ve (2013)

\bibitem[{{Anderson}(2014)}]{2014A&A...566L..10A}
{Anderson}, R.~I., \emph{{Tuning in on Cepheids: Radial velocity amplitude
  modulations. A source of systematic uncertainty for Baade-Wesselink
  distances}}, \emph{\aap} \textbf{566}, L10 (2014)

\bibitem[{{Anderson}(2016)}]{2016MNRAS.463.1707A}
{Anderson}, R.~I., \emph{{Discovery of Cycle-to-cycle Modulated Spectral Line
  Variability and Velocity Gradients in Long-period Cepheids}}, \emph{\mnras}
  \textbf{463}, 1707 (2016)

\bibitem[{{Anderson} et~al.(2015)}]{2015ApJ...804..144A}
{Anderson}, R.~I., et~al., \emph{{Revealing {$\delta$} Cephei's Secret
  Companion and Intriguing Past}}, \emph{\apj} \textbf{804}, 144 (2015)

\bibitem[{{Anderson} et~al.(2016{\natexlab{a}})}]{2016MNRAS.455.4231A}
{Anderson}, R.~I., et~al., \emph{{Investigating Cepheid {$\ell$} Carinae's
  cycle-to-cycle variations via contemporaneous velocimetry and
  interferometry}}, \emph{\mnras} \textbf{455}, 4231 (2016{\natexlab{a}})

\bibitem[{{Anderson} et~al.(2016{\natexlab{b}})}]{2016ApJS..226...18A}
{Anderson}, R.~I., et~al., \emph{{Vetting Galactic Leavitt Law Calibrators
  Using Radial Velocities: On the Variability, Binarity, and Possible Parallax
  Error of 19 Long-period Cepheids}}, \emph{\apjs} \textbf{226}, 18
  (2016{\natexlab{b}})

\bibitem[{{Baade}(1926)}]{1926AN....228..359B}
{Baade}, W., \emph{{{\"U}ber eine M{\"o}glichkeit, die Pulsationstheorie der
  {$\delta$} Cephei-Ver{\"a}nderlichen zu pr{\"u}fen}}, \emph{Astronomische
  Nachrichten} \textbf{228}, 359 (1926)

\bibitem[{{Baranne} et~al.(1996)}]{1996A&AS..119..373B}
{Baranne}, A., et~al., \emph{{ELODIE: A spectrograph for accurate radial
  velocity measurements.}}, \emph{\aaps} \textbf{119}, 373 (1996)

\bibitem[{{Barnes} et~al.(1988){Barnes}, {Moffett}, \&
  {Slovak}}]{1988ApJS...66...43B}
{Barnes}, T.~G., III, {Moffett}, T.~J., {Slovak}, M.~H., \emph{{Observational
  studies of Cepheids. VII - Radial velocities of faint Cepheids}},
  \emph{\apjs} \textbf{66}, 43 (1988)

\bibitem[{{Belopolsky}(1895)}]{1895ApJ.....1..160B}
{Belopolsky}, A., \emph{{The spectrum of delta Cephei.}}, \emph{\apj}
  \textbf{1}, 160 (1895)

\bibitem[{{Breitfelder} et~al.(2016)}]{2016A&A...587A.117B}
{Breitfelder}, J., et~al., \emph{{Observational calibration of the projection
  factor of Cepheids. II. Application to 9 Cepheids with HST/FGS parallax
  measurements}}, \emph{\aap} \textbf{587}, A117 (2016)

\bibitem[{{Britavskiy} et~al.(2018)}]{2017arXiv171105728B}
{Britavskiy}, N., et~al., \emph{{A new method of measuring centre-of-mass
  velocities of radially pulsating stars from high-resolution spectroscopy}},
  \emph{\mnras} \textbf{474}, 3344 (2018), \eprint{1711.05728}

\bibitem[{{Burki} et~al.(1982){Burki}, {Mayor}, \&
  {Benz}}]{1982A&A...109..258B}
{Burki}, G., {Mayor}, M., {Benz}, W., \emph{{The peculiar classical Cepheid HR
  7308}}, \emph{\aap} \textbf{109}, 258 (1982)

\bibitem[{{Butler}(1993)}]{1993ApJ...415..323B}
{Butler}, R.~P., \emph{{Cepheid velocity curves from lines of different
  excitation and ionization. I - Observations}}, \emph{\apj} \textbf{415}, 323
  (1993)

\bibitem[{{Casertano} et~al.(2016)}]{2016ApJ...825...11C}
{Casertano}, S., et~al., \emph{{Parallax of Galactic Cepheids from Spatially
  Scanning the Wide Field Camera 3 on the Hubble Space Telescope: The Case of
  SS CMa}}, \emph{\apj} \textbf{825}, 11 (2016)

\bibitem[{{Coulson} \& {Caldwell}(1985)}]{1985SAAOC...9....5C}
{Coulson}, I.~M., {Caldwell}, J.~A.~R., \emph{{Photometry and radial velocities
  of 27 southern galactic Cepheids}}, \emph{South African Astronomical
  Observatory Circular} \textbf{9}, 5 (1985)

\bibitem[{{Evans} et~al.(2015{\natexlab{a}})}]{2015AJ....150...13E}
{Evans}, N.~R., et~al., \emph{{Binary Properties from Cepheid Radial
  Velocities}}, \emph{\aj} \textbf{150}, 13 (2015{\natexlab{a}})

\bibitem[{{Evans} et~al.(2015{\natexlab{b}})}]{2015MNRAS.446.4008E}
{Evans}, N.~R., et~al., \emph{{Observations of Cepheids with the MOST
  satellite: contrast between pulsation modes}}, \emph{\mnras} \textbf{446},
  4008 (2015{\natexlab{b}})

\bibitem[{{Gaia Collaboration} et~al.(2016){Gaia Collaboration}, {Prusti}, {de
  Bruijne}, \& et~al.}]{2016A&A...595A...1G}
{Gaia Collaboration}, {Prusti}, T., {de Bruijne}, J.~H.~J., et~al., \emph{{The
  Gaia mission}}, \emph{\aap} \textbf{595}, A1 (2016)

\bibitem[{{Gorynya} et~al.(2002)}]{2002yCat.3229....0G}
{Gorynya}, N.~A., et~al., \emph{{Radial Velocities of Cepheids (Gorynya+
  1992-98)}}, \emph{VizieR Online Data Catalog} \textbf{III/229} (2002)

\bibitem[{{Imbert}(1999)}]{1999A&AS..140...79I}
{Imbert}, M., \emph{{D{\'e}termination des rayons de C{\'e}ph{\'e}ides. V.
  Vitesses radiales et dimensions de 22 C{\'e}ph{\'e}ides galactiques.}},
  \emph{\aaps} \textbf{140}, 79 (1999)

\bibitem[{{Joy}(1937)}]{1937ApJ....86..363J}
{Joy}, A.~H., \emph{{Radial Velocities of Cepheid Variable Stars}}, \emph{\apj}
  \textbf{86}, 363 (1937)

\bibitem[{{Kovacs} et~al.(1990){Kovacs}, {Kisvarsanyi}, \&
  {Buchler}}]{1990ApJ...351..606K}
{Kovacs}, G., {Kisvarsanyi}, E.~G., {Buchler}, J.~R., \emph{{Cepheid radial
  velocity curves revisited}}, \emph{\apj} \textbf{351}, 606 (1990)

\bibitem[{{Kraft}(1956)}]{1956PASP...68..137K}
{Kraft}, R.~P., \emph{{Double Lines in the Spectrum of the Classical Cepheid X
  Cygni}}, \emph{\pasp} \textbf{68}, 137 (1956)

\bibitem[{{Lindemann}(1918)}]{1918MNRAS..78..639L}
{Lindemann}, F.~A., \emph{{Note on the pulsation theory of Cepheid variables}},
  \emph{\mnras} \textbf{78}, 639 (1918)

\bibitem[{{Moore}(1929)}]{1929PASP...41...56M}
{Moore}, J.~H., \emph{{Note on the Long-Period System of Polaris}},
  \emph{\pasp} \textbf{41}, 56 (1929)

\bibitem[{{Nardetto} et~al.(2006)}]{2006A&A...453..309N}
{Nardetto}, N., et~al., \emph{{High resolution spectroscopy for Cepheids
  distance determination. I. Line asymmetry}}, \emph{\aap} \textbf{453}, 309
  (2006)

\bibitem[{{Nardetto} et~al.(2007)}]{2007A&A...471..661N}
{Nardetto}, N., et~al., \emph{{High-resolution spectroscopy for Cepheids
  distance determination. II. A period-projection factor relation}},
  \emph{\aap} \textbf{471}, 661 (2007)

\bibitem[{{Nardetto} et~al.(2009)}]{2009A&A...502..951N}
{Nardetto}, N., et~al., \emph{{High-resolution spectroscopy for Cepheids
  distance determination. V. Impact of the cross-correlation method on the
  p-factor and the {$\gamma$}-velocities}}, \emph{\aap} \textbf{502}, 951
  (2009)

\bibitem[{{Pepe} et~al.(2002)}]{2002A&A...388..632P}
{Pepe}, F., et~al., \emph{{The CORALIE survey for southern extra-solar planets
  VII. Two short-period Saturnian companions to HD 108147 and HD 168746}},
  \emph{\aap} \textbf{388}, 632 (2002)

\bibitem[{{Petterson} et~al.(2005){Petterson}, {Cottrell}, {Albrow}, \&
  {Fokin}}]{2005MNRAS.362.1167P}
{Petterson}, O.~K.~L., {Cottrell}, P.~L., {Albrow}, M.~D., {Fokin}, A.,
  \emph{{A spectroscopic study of bright southern Cepheids - a high-resolution
  view of Cepheid atmospheres}}, \emph{\mnras} \textbf{362}, 1167 (2005)

\bibitem[{{Pilecki} et~al.(2017)}]{2017ApJ...842..110P}
{Pilecki}, B., et~al., \emph{{Mass and p-factor of the Type II Cepheid
  OGLE-LMC-T2CEP-098 in a Binary System}}, \emph{\apj} \textbf{842}, 110 (2017)

\bibitem[{{Pont} et~al.(1994){Pont}, {Burki}, \& {Mayor}}]{1994A&AS..105..165P}
{Pont}, F., {Burki}, G., {Mayor}, M., \emph{{New radial velocities for 96 faint
  southern Cepheids}}, \emph{\aaps} \textbf{105}, 165 (1994)

\bibitem[{{Queloz} et~al.(2001)}]{2001A&A...379..279Q}
{Queloz}, D., et~al., \emph{{No planet for HD 166435}}, \emph{\aap}
  \textbf{379}, 279 (2001)

\bibitem[{{Rucinski}(1992)}]{1992AJ....104.1968R}
{Rucinski}, S.~M., \emph{{Spectral-line broadening functions of WUMa-type
  binaries. I - AW UMa}}, \emph{\aj} \textbf{104}, 1968 (1992)

\bibitem[{{Sanford}(1928)}]{1928ApJ....67..319S}
{Sanford}, R.~F., \emph{{On the period and radial velocity of the cluster-type
  variable RR Lyrae.}}, \emph{\apj} \textbf{67} (1928)

\bibitem[{{Sanford}(1949{\natexlab{a}})}]{1949PASP...61..135S}
{Sanford}, R.~F., \emph{{Double Absorption Lines in the Spectrum of W Vir}},
  \emph{\pasp} \textbf{61}, 135 (1949{\natexlab{a}})

\bibitem[{{Sanford}(1949{\natexlab{b}})}]{1949ApJ...109..208S}
{Sanford}, R.~F., \emph{{Radial Velocities of RR Lyr from Coud{\'e}
  Spectrograms}}, \emph{\apj} \textbf{109}, 208 (1949{\natexlab{b}})

\bibitem[{{Sasselov} \& {Karovska}(1994)}]{1994ApJ...432..367S}
{Sasselov}, D., {Karovska}, M., \emph{{On Cepheid diameter and distance
  measurement}}, \emph{\apj} \textbf{432}, 367 (1994)

\bibitem[{{Sasselov} et~al.(1989){Sasselov}, {Fieldus}, \&
  {Lester}}]{1989ApJ...337L..29S}
{Sasselov}, D.~D., {Fieldus}, M.~S., {Lester}, J.~B., \emph{{Infrared
  spectroscopy of Cepheids - Peculiar velocity structure and its effect on
  radii and distances}}, \emph{\apjl} \textbf{337}, L29 (1989)

\bibitem[{{Sasselov} \& {Lester}(1990)}]{1990ApJ...362..333S}
{Sasselov}, D.~D., {Lester}, J.~B., \emph{{Infrared spectroscopy of Cepheids.
  II - Line profiles from different atmospheric layers}}, \emph{\apj}
  \textbf{362}, 333 (1990)

\bibitem[{{Shapley}(1916)}]{1916ApJ....44..273S}
{Shapley}, H., \emph{{The variations in spectral type of twenty Cepheid
  variables}}, \emph{\apj} \textbf{44}, 273 (1916)

\bibitem[{{Stibbs}(1955)}]{1955MNRAS.115..363S}
{Stibbs}, D.~W.~N., \emph{{Radial velocities of cepheid variable stars in the
  southern hemisphere}}, \emph{\mnras} \textbf{115}, 363 (1955)

\bibitem[{{Szabados}(2003)}]{2003IBVS.5394....1S}
{Szabados}, L., \emph{{Database on Binaries among Galactic Classical
  Cepheids}}, \emph{IBVS} \textbf{5394} (2003)

\bibitem[{{Wallerstein}(1972)}]{1972PASP...84..656W}
{Wallerstein}, G., \emph{{Motions in the Outer Layers of the 27-Day Cepheid T
  Monocerotis}}, \emph{\pasp} \textbf{84}, 656 (1972)

\bibitem[{{Wallerstein} et~al.(1992)}]{1992MNRAS.259..474W}
{Wallerstein}, G., et~al., \emph{{Metallic-line and H-alpha radial velocities
  of seven southern Cepheids - A comparative analysis}}, \emph{\mnras}
  \textbf{259}, 474 (1992)

\bibitem[{{Wesselink}(1946)}]{1946BAN....10...91W}
{Wesselink}, A.~J., \emph{{The observations of brightness, colour and radial
  velocity of {$\delta$} Cephei and the pulsation hypothesis)}}, \emph{\bain}
  \textbf{10}, 91 (1946)

\end{thebibliography}

\end{document}